\documentstyle[prl,aps,epsfig,twocolumn]{revtex}

\draft

\newlength{\textwidthm}
\begin{document}
\title{Manipulating Light Pulses via Dynamically controlled Photonic 
Bandgap} 
\author{A. Andr\'{e} and M. D. Lukin}
\address{Physics Department and ITAMP, Harvard University, 
Cambridge, Massachusetts 02138}
\date{\today}
\maketitle
\begin{abstract}
When a resonance associated with electromagnetically induced
transparency (EIT) in an atomic ensemble is modulated by an off-resonant
standing light wave,  a band of frequencies can appear for
which light propagation is forbidden. We show that dynamic control of such 
a bandgap can be 
used to coherently convert a propagating light pulse into a stationary 
excitation with non-vanishing photonic component. This can be accomplished 
with  high efficiency and negligble noise even at a level of few-photon
quantum fields thereby facilitating possible applications in 
quantum nonlinear optics and quantum information.
\end{abstract}



Techniques for coherent control of light-matter interaction
are now actively explored for storing and manupilating
quantum states of photons. In particular, using electromagnetically 
induced transparency (EIT) \cite{EIT1,EIT2} and adiabatic following
of ``dark-state polaritons'' \cite{polaritons}, the group
velocity of light pulses can be dramatically decelerated and their 
quantum state can be mapped onto metastable collective states of 
atomic ensembles \cite{storage}. 

In contrast to such a coherent absorption 
process, the present Letter describes how a propagating 
light pulse can be converted  into a stationary  excitation with 
non-vanishing photonic component. 
This is accomplished via controlled modification of the photonic density
of states in EIT media by modulating the refractive index with an
off-resonant standing light wave. 
By varying the properties of the resulting photonic band structure 
in time, the original light pulse can be converted into an excitation 
inside the bandgap where its propagation is forbidden. Long storage of
excitations
with non-vanishing photonic component may open interesting prospects for
enhancement of nonlinear optical interactions \cite{nlo1,nlo3}. 
In particlular, an 
intriguing and practically important \cite{Q-comp,Q-info} 
application of this effect for interactions between few-photon
fields is dicussed in the concluding paragraph of this Letter.

Before proceeding, we note that there exists a substantial litterature on
photonic bandgap \cite{PBG} materials.  Recently photonic bandgap structures
have been investigated theoretically
\cite{mabuchi} for strong coupling of single
atoms with photons. Photonic bandgap based on interaction with atoms in an
optical lattice were also investigated \cite{deutsch}. We also note 
other related work on EIT-based
control of the propagation properties of light in atomic media \cite{EIT-prop}.

The key idea of the present approach can be qualitatively understood 
by first considering a medium consisting of stationary atoms with a level
structure shown in Fig. 1a. The atoms are interacting with a weak signal
field and two strong fields.
The running wave control field $\Omega_c$ 
is tuned to resonant frequency  of the $|b\rangle\rightarrow|c\rangle$ 
transition. In the absence of the field $\Omega_s$, this situation corresponds
to the usual EIT: in the vicinity of a frequency corresponding to 
two-photon resonance the medium becomes transparent for a signal
field. This transparency is accompanied by a steep variation of
the refractive index.

The dispresion relation can be further manipulated by applying an off-resonant
standing wave field with Rabi frequency $\Omega_s(z)=2\Omega_s\cos(k_sz)$
and a frequency detuning $\Delta$. 
This field induces an effective shift of resonant frequency (light shift)
that varies periodically in space, resulting in a spatial modulation 
of the index of refraction according to $\delta n(z) = (c/v_g) 
4\frac{\Omega_s^2}{\Delta}\cos^2(k_sz)$, where $c/v_g$ is the ratio of
speed of
light in vacuum to group velocity in the medium.
When the modulation depth is sufficiently large, 
signal light propagating near atomic resonance in the forward $z$
direction  with wavenumber $k$ near $k_s$ may undergo Bragg scattering
into the backward propagating mode with wavenumber $-k$. In direct analogy
to e.g., optical interferometers, the scattering of the counterpropagating
fields into each other can modify the photonic density of states. 
In particular,  
a range of frequencies (``photonic bandgap'') can appear 
for which light propagation is forbidden \cite{yariv}. According to 
a standard technique to analyze the resulting band structure,
Bloch's theorem can be applied so that the propagating solutions
obey $E(z+a)=e^{i K a}E(z)$, where $K$ is the Bloch wave vector.
Imposing this condition and assuming that the wave vectors of the fields 
are close  ($k\simeq k_s$), we can solve for the band structure and obtain
near two-photon resonance
\begin{equation}
\cos(Ka)=\cosh\left(\frac{g^2N}{\Omega_c^2}a\sqrt{\Delta_s^2-
(\omega-\omega_{ba})^2}\right),
\end{equation}
where $g=\wp\sqrt{\frac{\nu}{2\hbar\epsilon_0 V}}$ is the atom-field
coupling constant, $N$ is the number of atoms,
$\Delta_s=\Omega_s^2/\Delta$ is the amplitude of the
light shift modulation, $\wp$ is the dipole moment of the $a-b$
transition, $V$ the quantization volume and the factor $g^2N/\Omega_c^2$
corresponds to $c/v_g$. 
For frequencies such that $|\omega-\omega_{ba}|\leq|\Delta_s|$ 
a {\it bandgap} is created: the Bloch wavevector
acquires an imaginary part and the propagation of waves in the medium is
forbidden. For an outside observer such a medium can be viewed as a 
mirror: an incident wave with frequency inside the
bandgap would undergo almost perfect reflection. Calculations in 
Fig. 2 indicate that this qualitative result remains valid even 
for realistic EIT conditions, including a finite transparency 
bandwidth and finite ground-state decoherence rate $\gamma_{bc}$.

A specific, distinguishing feature of the present scheme is
the possibility of {\it dynamically} changing the
properties of the medium by switching in time the fields $\Omega_s(t)$
and $\Omega_c(t)$ on and off. 
In particular, by combining the techniques of \cite{storage} with the
present idea, the following scenario can be implemented: first, with the
standing wave turned off $\Omega_s=0$, a forward propagating pulse is
stored in the medium as a Raman coherence
between levels $|b\rangle$ and $|c\rangle$. Then, by switching on both the
``control'' field $\Omega_c$ and the standing wave field $\Omega_s$,
the pulse can be released into the bandgap and initially propagates in
the forward direction.
In the presence of a bandgap, the forward
($\hat{{\mathcal E}}_+$) and backward ($\hat{{\mathcal E}}_-$) components 
are coupled due to Bragg scattering off the index grating,
so that amplitude in the forward mode $+k$ is
converted into amplitude in the backward mode $-k$ and vice versa.
In this case, the pulse can be effectively trapped in the photonic
bandgap.

We now turn to a detailed description of the dynamic trapping procedure.
We are interested here in the propagation of fields with
possibly non-trivial statistics, such as single photon fields, so that a
quantum description is used.
In the presence of the standing wave it is convenient to decompose the 
propagating signal fields into two  slowly varying components 
$\hat{{\mathcal E}}_+$ (propagating forward) and
$\hat{{\mathcal E}}_-$ (propagating backward) so that the electric field
is
$\hat{E}(z,t) = \sqrt{\frac{\hbar \nu}{2\epsilon_0
V}}\left[\sum_{\sigma=\pm}\hat{{\mathcal
E}}_\sigma(z,t)e^{i(\nu/c)(\sigma z-ct)}+{\rm h.c.}\right]$
and couples resonantly to the transition between the ground state
$|a\rangle$ and the excited state $|b\rangle$, the carrier frequency of
the optical field being $\nu=\omega_{ab}$. 
Two time dependent classical driving fields with Rabi-frequencies 
$\Omega_s(t)$ and $\Omega_c(t)$ are used to control the propagation as
shown if Fig. 1a.

To describe the quantum properties of the medium, we use collective slowly 
varying atomic operators \cite{fleisch1} 
$\hat{\sigma}_{\mu\nu}(z,t)=\frac{1}{N_z}\sum_{j=1}^{N_z}|\mu_j\rangle
\langle\nu_j|e^{-i\omega_{\mu\nu}t}$
where the sum is performed over a small but macroscopic volume containing
$N_Z\gg 1$ atoms around position $z$.
The interaction hamiltonian is then, in a rotating frame
\begin{eqnarray}
\hat{H}&=&\frac{N}{L}\int
dz\left\{\Delta\hat{\sigma}_{dd}-\left[g(\hat{{\mathcal E}}_+e^{ik_0z}
+\hat{{\mathcal E}}_-e^{-ik_0z})\hat{\sigma}_{ab}\right.\right. \nonumber
\\
&+& \left.\left. \Omega_c e^{ik_cz}\hat{\sigma}_{ac}
+2\Omega_s \cos(k_sz)\hat{\sigma}_{dc}+{\rm h.c.}\right]\right\},
\end{eqnarray}
where $k_0=\nu/c$ and $N$ the number of atoms. 

Since the two propagating fields $\hat{{\mathcal E}}_+$ and
$\hat{{\mathcal E}}_-$
interact with the atoms, we expect an optical coherence
$\hat{\sigma}_{ba}$ to
appear as well as a Raman coherence $\hat{\sigma}_{bc}$. Moreover these
two
fields will give rise to coherences with distinct spatial variations,
i.e., varying as $e^{ik_0z}$ for the component of 
$\hat{\sigma}_{ba}$ induced by $\hat{{\mathcal E}}_+$ 
while that due to 
$\hat{{\mathcal E}}_-$ will vary as
$e^{-ik_0z}$. 
We therefore decompose the optical and Raman coherences according to these
two distinct spatial variations
$\hat{\sigma}_{ba}(z,t)=\hat{\sigma}_{ba}^+(z,t)e^{ik_0z}+
\hat{\sigma}_{ba}^-(z,t)e^{-ik_0z}$
and 
$\hat{\sigma}_{bc}(z,t)=\hat{\sigma}_{bc}^+(z,t)e^{i(k_0-k_c)z}+
\hat{\sigma}_{bc}^-(z,t)e^{-i(k_0+k_c)z}$. Using slowly varying envelopes, we 
then have the equations of motion for the forward and backward modes
\begin{eqnarray}
\left(\frac{\partial}{\partial t}\pm c\frac{\partial}{\partial
z}\right)\hat{{\mathcal E}}_\pm(z,t)=igN\hat{\sigma}_{ba}^\pm(z,t). 
\end{eqnarray}

Assuming weak quantum fields and solving perturbatively, we find to lowest
order in the weak fields and in an adiabatic approximation (assuming
$\Omega_s(t)$ and $\Omega_c(t)$ change in time slowly enough 
\cite{polaritons})
\begin{eqnarray}
\hat{\sigma}_{ba}^\pm(z,t)&=&-\frac{i}{\Omega_c}\left[\frac{\partial}{\partial
t}\hat{\sigma}_{bc}^\pm(z,t)-i\Delta_se^{\pm 2i\Delta kz}
\hat{\sigma}_{bc}^\mp(z,t)\right] \\ 
\hat{\sigma}_{bc}^\pm(z,t)&=&-\frac{g\hat{{\mathcal E}}_\pm(z,t)}{\Omega_c}
-i\frac{\hat{F}_{ba}^\pm(t)}{\Omega_c},
\end{eqnarray}
where $\Delta k=k_s-k_0$, $\Delta_s=|\Omega_s|^2/\Delta$ is the amplitude
of the spatially modulated light shift caused by the standing wave field
$\Omega_s$ and $\hat{F}_{ba}^\pm(t)$ are $\delta$-correlated noise
forces. 
Note that in the adiabatic limit the noise forces are
negligible \cite{polaritons}.
The propagation equations are thus 
\begin{eqnarray}
\left(\frac{\partial}{\partial t}\pm c\frac{\partial}{\partial
z}\right)
\hat{{\mathcal E}}_\pm(z,t)=
&-&\frac{g^2N}{\Omega_c}\frac{\partial}{\partial t}
\frac{\hat{{\mathcal E}}_\pm(z,t)}{\Omega_c}\nonumber 
\\
&+&i\frac{g^2N}{\Omega_c}\Delta_s
\frac{\hat{{\mathcal E}}_\mp(z,t)}{\Omega_c}e^{\pm 2i\Delta kz},
\end{eqnarray}
which indicates that the forward and backward slowly propagating
modes become coupled. Specifically, the first term on the right-hand side
gives rise to
propagation at the group velocity $v_g=c/(1+\frac{g^2N}{\Omega_c^2})$
\cite{slowlight}
while the second term gives rise to coupling between the forward and
backward propagating modes. 
This coupling is optimum when the effective phase
matching $\Delta k=k_s-k_0=0$ is achieved \cite{yariv}. 
Note that both the ``control'' field $\Omega_c$ and
the standing wave amplitude $\Omega_s$ can be time-dependent and that as
long as changes are slow enough (adiabatic limit \cite{polaritons}) the
above equations describe the correct dynamics of the coupled modes.

To obtain a solution in the case of time-dependent fields $\Omega_c(t)$
and $\Omega_s(t)$, we introduce new quantum
fields $\hat{\Psi}_+(z,t)$ and $\hat{\Psi}_-(z,t)$ (forward and backward
propagating dark-state polaritons \cite{polaritons}) 
$\hat{\Psi}_\pm(z,t)=\cos\theta(t)\hat{{\mathcal E}}_\pm(z,t)
-\sin\theta(t)\sqrt{N}\hat{\sigma}_{bc}^\pm(z,t)$,
where $\tan^2\theta(t)=\frac{g^2N}{\Omega_c(t)^2}$ is the mixing angle
between the photon and matter components of the polariton.
The polaritons then obey the coupled equations 
\begin{eqnarray}
\left(\frac{\partial}{\partial\tau}
\pm c\frac{\partial}{\partial z}\right)
\hat{\Psi}_\pm &=& i\Delta_s\tan^2\theta(t)\hat{\Psi}_\mp,
\label{propeqn}
\end{eqnarray}
where $\tau(t)=\int^{t}dt'\;\cos^2\theta(t')$.
Eq. (\ref{propeqn}) describes propagation with velocity
$v_g(t)=c\cos^2\theta(t)$ of
the two polaritons (traveling in opposite directions) and
coupling with rate $\Delta_s(t)\sin^2\theta(t)$. Note that in the limit
$c\gg v_g$, the photonic component
$\hat{{\mathcal E}}_\pm\simeq (\Omega_c/g\sqrt{N})\hat{\Psi}_\pm$ is
finite for non-zero control field $\Omega_c$.

We consider now the scenario in which the standing wave beams
are initially off and the control field is on, with Rabi frequency
$\Omega_c^{in}$ (corresponding  to a group velocity $v_g^{in}$). A forward 
propagating photon wavepacket can then be stored in the medium in the form
of a Raman coherence $\hat{\sigma}_{bc}^+(z,t)$ and subsequently released
\cite{storage}.
We consider the case when the standing wave field is first switched on, 
establishing the bandgap, followed by the control field (with Rabi
frequency $\Omega_c^0$ corresponding to a group velocity $v_g^0$, 
possibly different from $v_g^{in}$),
releasing the pulse in the bandgap medium. For simplicity we
consider the case when the standing wave is switched on before or
simultaneously with the control field, so that, 
the coupling rate $\Delta_s(\tau)\tan^2\theta(\tau)$ does not depend on
$\tau$.
In this case, we solve (\ref{propeqn}) by
Fourier transforming
$\hat{\Psi}_\pm(z)=\frac{1}{2\pi}\int{dk\;e^{ikz}\hat{\Psi}_\pm(k)}$ to
obtain
\begin{eqnarray}
\hat{\Psi}_+(k,\tau) &=&
\left[\cos(\zeta\tau)-i\frac{kc}{\zeta}\sin(\zeta\tau)\right]\hat{\Psi}_+(k,0) 
\nonumber \\
\hat{\Psi}_-(k,\tau) &=&
-i\frac{\chi}{\zeta}\sin(\zeta\tau)\hat{\Psi}_+(k,0),
\label{solrabi}
\end{eqnarray}
where $\chi\equiv\Delta_s(\tau)\tan^2\theta(\tau)$ and
$\zeta=\sqrt{(kc)^2+\chi^2}$. 
According to (\ref{solrabi}), the various 
Fourier components (wavenumber $k$) of the pulse cycle back and forth
between the corresponding forward and backward modes at a rate which
depends weakly on $k$. In particular
when the spatial extent of the pulse inside the medium
is large enough to give a relevant range of wavenumber negligible compared
to the strength of the coupling of forward and backward modes, pulse
distortion is negligible and the spatial envelopes
have the time dependence
$\hat{\Psi}_+(z,\tau)=\cos(\chi\tau)\hat{\Psi}_+(z,0)$
and$\hat{\Psi}_-(z,\tau)=-i\sin(\chi\tau)\hat{\Psi}_+(z,0)$.
The wavepacket periodically cycles between a forward and backward propagating
component, the result of which is {\it trapping} of the
pulse in the medium as shown in Fig. 3. 
The wavepacket is trapped as a combination of light
pulse and Raman coherence. 

The above analysis involves an adiabatic approximation and ignores the
decay of Raman coherence. 
In order to ignore the motion compared to the coupling we require
that $\chi\gg kc$ and since the maximum $k$ can be estimated from the
initial length of the pulse in the medium we find that this is equivalent
to requiring that $\Delta_sT\gg \frac{v_g^0}{v_g^{in}}$,
where $T$ is the duration of the initial pulse. 
As seen from (\ref{solrabi}) the effect
of non-zero values of $k$ is that the trapped pulse become spatially
distorted. Expanding $\sqrt{\chi^2+(kc)^2}\tau\simeq
[\chi+(kc)^2/(2\chi)]\tau$ we need $\tau\ll\chi/(kc)^2$, which gives after
expressing $\tau$ in terms of real time $t$ that
$\frac{t_{int}}{T}\ll\Delta_sT(v_g^{in}/v_g^0)^2$, where
$t_{int}$ is the
maximum time during which the pulse may be trapped without suffering
distortion. Furthermore, taking into
account the limits imposed by adiabaticity (i.e., modulation of index 
occurs within the transparency window $\Delta_s\ll(\Omega_c^0)^2/\gamma$)
and the
fact that the trapped pulse must fit inside the medium when travelling at 
the reduced group velocity, we find that the trapping or interaction time
is limited by
\begin{equation}
t_{int}\lesssim\frac{g^2N}{\gamma_{ab}(c/L)}\frac{v_g^{in}}{v_g^0}T
,\frac{1}{\gamma_{bc}},
\end{equation}
where $L$ is the length of the medium. This limiting quantity 
corresponds to the density length-product and can be rather large for 
optically dense medium.

To summarize, we have shown that by spatially modulating the dispersive
feature of the EIT resonance it is possible to induce a photonic bandgap.
By dynamically controlling the resulting band structure, a
propagating light pulse can be converted into a stationary excitation 
which is effectively trapped in the medium.

To conclude, we note some interesting avenues opened by this
work.
First, we note that the present work is not restricted to the
use of stationary or cold atoms, for example a Doppler-free 
configuration involving pairs of copropagating fields 
is shown in Fig. 1b. In this case, the two polaritons are asociated with
distinct atomic states $|c\rangle$ and $|c'\rangle$. Each polariton
corresponds to a Doppler-free Raman
configuration and they are coupled by a Doppler-free two-photon
transition.
Second, this work may open interesting prospects for nonlinear optics.
For example, a trapped photonic excitation can be used to induce a light
shift via interaction with another atom-like polariton.
Large nonlinear phase shifts at the single-photon level 
can be expected and open up the
way for possible applications in quantum non-linear optics and quantum
information without the limitations associated with traveling wave 
configurations 
\cite{nlo3,imamoglu} and without invoking cavity QED techniques \cite{cqed}.
Finally, it is intriguing to consider extension of these ideas to
manipulate photonic bandgap in condensed matter.

This work was supported by the NSF through ITR program and the grant to the 
ITAMP.

\def\etal{\textit{et al.}}


\begin{figure}[ht]
\centerline{\epsfig{file=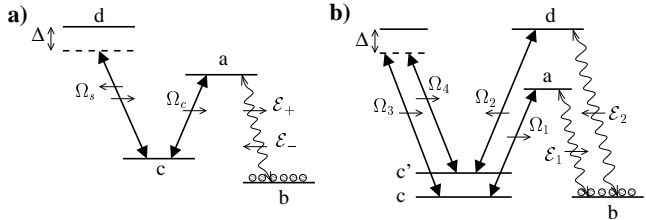,width=8.5cm}}
 \vspace*{2ex}
 \caption{Atomic level configuration for EIT-induced photonic bandgap.
a) Stationary atoms scheme, b) moving atoms scheme.
The standing wave of Rabi frequency
$\Omega_s$ is detuned by $\Delta$ from resonance with the
$|c\rangle\rightarrow |d\rangle$ transition, giving rise to a
spatially modulated light shift $\Delta_s(z,t)=|\Omega_s(z,t)|^2/\Delta$.
} 
\end{figure}



\begin{figure}[ht]
\centerline{\epsfig{file=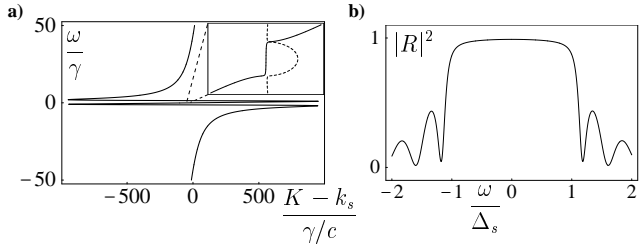,width=8.5cm}}
 \vspace*{2ex}
 \caption{a) Dispersion relation $\omega(K)$ for waves propagating in
EIT-induced bandgap medium. Plotted is the detuning from two-photon
resonance $\omega=\nu-\nu_0$ in units of linewidth $\gamma$ vs. the Bloch
wavevector $K$ in units of $\gamma/c$. 
For $|\omega|\gg\gamma$, the dispersion relation
corresponds to the usual $\nu=c K$ while near resonance the slope is
changed to the group velocity $\nu=v_g K$. For $|\omega|\leq\Delta_s$
a bandgap appears, as shown in the inset 
(full line: real part of $K-k_s$, dashed line: imaginary part). 
b) Reflection coefficient
vs. frequency near EIT resonance for a sample of $^{87}{\rm Rb}$
atoms corresponding to a density $n\sim 10^{13}{\rm
cm}^{-3}$ and length of medium $L\sim 4{\rm cm}$ 
($\gamma_{bc}=1{\rm kHz}$,
$\gamma\sim 10{\rm MHz}$, $g\sqrt{N}\sim 400{\rm MHz}$, $c/v_g\sim
10^3$ and $\Delta_s\sim 400 {\rm kHz}$), giving a peak reflectivity
of $98.9\%$.} \end{figure}



\begin{figure}[ht]
\centerline{\epsfig{file=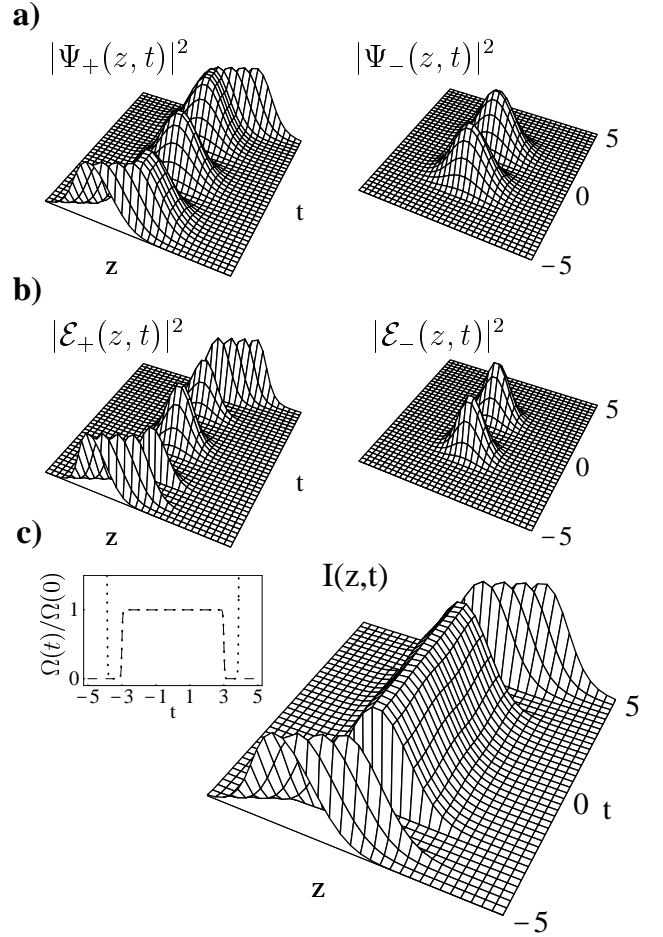,width=8.5cm}}
 \vspace*{2ex}
 \caption{a) Amplitude of
forward and backward propagating polaritons $\Psi_+(z,t)$ and $\Psi_-(z,t)$. 
b) Corresponding electric fields and c) total
intensity (forward and backward components) averaged over the optical
wavelength. Also shown is the time-dependence of the
``control'' field $\Omega_c(t)$ (dotted line) and of standing wave field
$\Omega_s(t)$ (dashed line). 
Note that $v_g^{in}/v_g^0\sim 15$ here so that
initial motion of the pulse is noticeable on these plots. 
Axes are in arbitrary units.} \end{figure}


\end{document}